\documentclass{llncs}
\usepackage{graphicx}
\usepackage{amsfonts}
\usepackage{amssymb}
\usepackage{latexsym}
\newcommand{\full}[1]{#1}

\newcommand{\conf}[1]{}
\newcommand{\squeezelist}{\setlength{\itemsep}{0pt}}

\def\b{{\beta}}
\def\a{{\alpha}}

\def\l{{\lambda}}
\def\bP{{\partial P}}
\def\bQ{{\partial Q}}
\def\T{{\cal T}}
\newenvironment{pf}{\unskip{\bf Proof:}}{\unskip{\hfill $\Box$}}

\newcommand{\lemlab}[1]{\label{lemma:#1}}
\newcommand{\theolab}[1]{\label{theo:#1}}

\newcommand{\figlab}[1]{\label{fig:#1}}
\newcommand{\seclab}[1]{\label{section:#1}}

\newcommand{\lemref}[1]{\ref{lemma:#1}}

\newcommand{\theoref}[1]{\ref{theo:#1}}

\newcommand{\figref}[1]{\ref{fig:#1}}
\newcommand{\eqref}[1]{(\ref{eq:#1})}

{\catcode`\@=11
\gdef\setft#1#2#3{%
\def\@oddfoot{
{\setbox0=\hbox{#1}
\setbox1=\hbox{#3}
\ifdim\wd0>\wd1
\dimen0=\wd0
\box0\hfil#2\hfil\hbox to\dimen0{\hfil\hfil\box1}
\else \dimen0=\wd1
\hbox to\dimen0{\box0\hfil }\hfil#2\hfil\box1 \fi
}}} }

\def\complaint#1{}
\def\withcomplaints{
\newcounter{mycomplaints}
\def\complaint##1{\refstepcounter{mycomplaints}%
\ifhmode%
\unskip%
{\dimen1=\baselineskip \divide\dimen1 by 2 %
\raise\dimen1\llap{\tiny -\themycomplaints-}}\fi%
\marginpar{\tiny [\themycomplaints]: ##1}}%
}

\conf{
\title{{\bf Enumerating Foldings and Unfoldings\\between Polygons and Polytopes}\\
{\em (Abstract)}}
\date{}
}%conf
\full{
\title{Enumerating Foldings and Unfoldings\\between Polygons and Polytopes}
\titlerunning{Enumerating Foldings and Unfoldings}
\author{%
Erik D. Demaine\inst{1}
\and 
Martin L. Demaine\inst{1}
\and 
Anna Lubiw\inst{1}
\and
Joseph~O'Rourke\inst{2}
}
\authorrunning{Demaine, Demaine, Lubiw, O'Rourke}
\institute{
Dept.\ Comput.\ Sci., Univ. of Waterloo,
Waterloo, ON N2L 3G1, Canada.
\email{\{eddemaine,mldemaine,alubiw\}@uwaterloo.ca}.
\and
Dept.\ Comput.\ Sci., Smith Col\-lege, North\-ampton,
MA 01063, USA.
\email{orourke@cs.smith.edu}.
}
}%full

\conf{
\author{%
Erik D. Demaine\thanks{
Dept.\ Comput.\ Sci., Univ. of Waterloo,
Waterloo, ON N2L 3G1, Canada.
{\tt \{eddemaine,\-mldemaine,\-alubiw\}@uwaterloo.ca}.
}%thanks
\and
Martin L. Demaine\footnotemark[1]
\and
Anna Lubiw\footnotemark[1]
\and
Joseph~O'Rourke\thanks{Dept.\ Comput.\ Sci., Smith Col\-lege, North\-ampton,
MA 01063, USA.
{\tt orourke@cs.smith.edu}.
}%thanks
}%author
}%conf

\begin{document}
\maketitle
\begin{abstract}
We pose and answer several questions concerning the number
of ways to fold a polygon to a polytope, and how many polytopes
can be obtained from one polygon; and the analogous questions
for unfolding polytopes to polygons.  Our answers are, roughly:
exponentially many, or nondenumerably infinite.
\end{abstract}

%%%%%%%%%%%%%%%%%
%\newpage
%\tableofcontents
%\newpage
%\pagenumbering{arabic}
%\setcounter{page}{1}
%%%%%%%%%%%%%%%%%

\section{Introduction}
\seclab{Introduction}
We explore the process of folding a simple polygon
by gluing its perimeter shut to form
a convex polyhedron, and its reverse, cutting a
convex polyhedron open and flattening its surface to a
simple polygon.
We restrict attention to convex polyhedra (henceforth,
{\em polytopes\/}),
and to simple (i.e., nonself-intersecting,
nonoverlapping) polygons (henceforth, {\em polygons}).
The restriction to nonoverlapping polygons is natural,
as this is important to manufacturing
applications~\cite{o-fucg-00}.
The restriction to convex polyhedra is made primarily to
reduce the scope of the problem.
See ~\cite{bddloorw-uscop-98} and~\cite{bdek-up-99}
for a start on unfolding nonconvex polyhedra.

We enumerate foldings and unfoldings based on two criteria
of indistinguishability: 
geometric congruence, and 
combinatoric equivalence.  The latter especially will need
further specification to become precise, but to presage our
results crudely, we show that both the number of foldings
and the number of unfoldings can be exponential in
the number of vertices $n$ of the polygon/polytope.
Similarly, we show that 
polygons may fold and polytopes unfold to
an infinite number of incongruent polytopes/polygons.
We obtain sharper results when attention is restricted to
convex polygons.
Proofs and details not provided here may be found in~\cite{ddlo-ecerfu-00}.

We will use $P$ throughout the paper for a polygon,
and $Q$ for a polytope, 
$\bP$ and $\bQ$ respectively for their boundaries,
and $n$ for the number of their vertices.

\section{Aleksandrov's Theorem}
\seclab{Aleksandrov}
A key tool in our work is
a far-reaching generalization of
Cauchy's rigidity theorem
proved by Aleksandrov~\cite{a-kp-58} 
that gives simple conditions for any folding
to a polytope.
%Let $P$ be a polygon and $\bP$ its boundary.
\full{
A {\em gluing\/}
maps $\bP$ to $\bP$ in a length-preserving
manner, as follows.
$\bP$ is partitioned by a finite number of distinct points
into a collection of open intervals whose closure covers $\bP$.
Each interval is mapped one-to-one (i.e., {\em glued\/})
to another interval of equal length.
Corresponding endpoints of glued intervals are glued
together (i.e., identified).
Finally, gluing is considered
transitive: if points $a$ and $b$ glue to point $c$, then
$a$ glues to $b$.
}%full
\conf{
A {\em gluing\/}
maps $\bP$ to $\bP$ in a length-preserving
manner.
}%conf
Aleksandrov proved that any gluing that satisfies these two conditions
corresponds to a unique polytope:
\begin{enumerate}
\squeezelist
%\item The gluing uses up the whole perimeter, i.e.,
%the union of the glued intervals and endpoints is $\bP$.
\item No more than $2 \pi$ total face angle is glued together at any point; and
\item The complex resulting from the gluing is homeomorphic to a sphere.
\end{enumerate}
%Aleksandrov calls any complex (not necessarily a single polygon)
%that satisfies these properties a {\em net}~\cite{a-kp-58}.
We call a gluing that satisfies these conditions
an {\em Aleksandrov gluing}.
Although an
Aleksandrov gluing of a polygon forms a unique polytope,
it is an open problem to compute
the three-dimensional structure of the polytope~\cite{o-fucg-00}.
Henceforth we will say a polygon {\em folds\/} to
a polytope whenever it has an Aleksandrov gluing.

\full{
We should mention two features of Aleksandrov's theorem.
First, the polytope whose existence is guaranteed may be
{\em flat}, that is, a doubly-covered convex polygon.
We use the term ``polytope'' to include flat polyhedra.
Second,
condition~(2) specifies a face angle $\le 2 \pi$.
The case of equality with $2 \pi$ leads to a point
on the polytope at which there is no curvature,
i.e., a nonvertex.  We make explicit
what counts as a vertex below.
}%full

\section{Geometrical Congruence}
\seclab{Geometrical.Congruence}
In this section we address these two natural questions:
\begin{enumerate}
\squeezelist
\item How many geometrically different polytopes may be folded from one polygon?
\item How many geometrically different polygons may be unfolded from one polytope?
\end{enumerate}
Here ``geometrically different'' means incongruent.
Although we mentioned the rough answer to both question is `infinite,'
there are several nuances in the details.  For example, the
answer to the first question is: `sometimes infinite,' 
whereas the answer to the second is: `always infinite.'

\subsection{Congruence: Folding}
\seclab{Congruence.Folding}
We start with a natural and easily proved claim:

\begin{lemma}
Some polygons cannot be folded to any polytope.
\lemlab{unf}
\end{lemma}
\full{
An example is shown in Fig.~\figref{unf}.

It is natural to wonder what the chances are that a random polygon
could fold to a polytope.
This is difficult to answer without a precise definition of
``random,''
but we feel any reasonable definition would
lead to the same answer:
{\sc zero}.
We provide support for this conjecture in~\cite{ddlo-ecerfu-00}.
Despite this evidence for the rare ability to fold, 
convex polygons are fertile in their folding,
as we now demonstrate.
}%full

Let $x \in \bP$ be an arbitrary point on the boundary of $P$,
and let $y \in \bP$ be the midpoint of perimeter $L$ around $\bP$
measured from $x$.
Let $(x,y)$ be the open interval
of $\bP$ counterclockwise from $x$ to $y$.
Define a {\em perimeter-halving gluing\/}
as one which glues $(x,y)$ to $(y,x)$.
A consequence of Aleksandrov's theorem is:

\full{
See Fig.~\figref{perim.halving} for an example.
%%%%%%%%%%%%%%%%%%%%%%%%%%%%%%%%%Figure Begin
\begin{figure}[htbp]
\begin{minipage}[t]{0.45\linewidth}
\centering
\includegraphics[height=3.25cm]{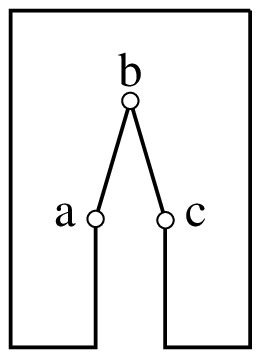}
\caption{An unfoldable polygon.}
\figlab{unf}
\end{minipage}%
\hspace{0.10\linewidth}
%\end{figure}
%%%%%%%%%%%%%%%%%%%%%%%%%%%%%%%%%Figure End
%%%%%%%%%%%%%%%%%%%%%%%%%%%%%%%%%Figure Begin
%\begin{figure}[htbp]
\begin{minipage}[t]{0.45\linewidth}
\centering
\includegraphics[width=5cm]{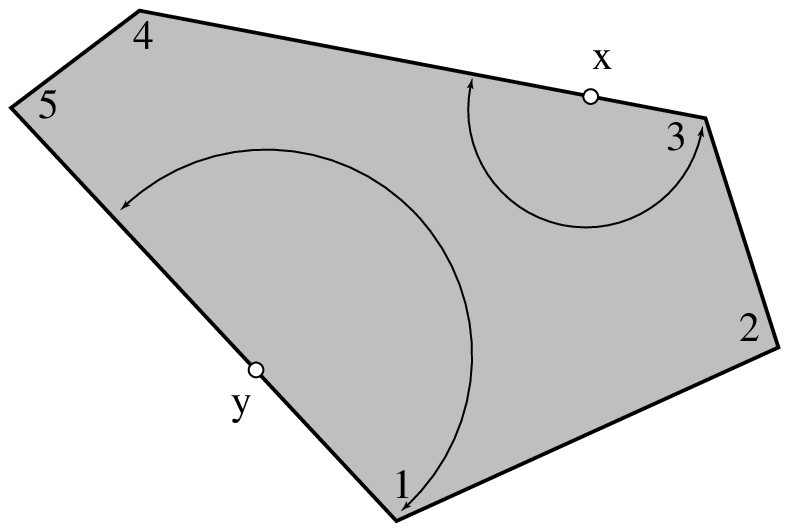}
\caption{A perimeter-halving fold of a pentagon. The gluing mappings
of vertices $v_1$ and $v_3$ are shown.}
\figlab{perim.halving}
\end{minipage}
\end{figure}
%%%%%%%%%%%%%%%%%%%%%%%%%%%%%%%%%Figure End
}%full

\begin{lemma}
Every convex polygon folds to a polytope via perimeter halving
for every $x \in \bP$.
\lemlab{perim.halving}
\end{lemma}

\noindent
This result can be strengthened:
\begin{theorem}
Any convex polygon $P$ folds, via perimeter halving,
to a nondenumerably infinite number of
noncongruent polytopes.
\theolab{continuum}
\end{theorem}

Using a different type of folding, we can show
that every rectangle folds to a continuum of tetrahedra.
See Fig.~\figref{tetra}.
%%%%%%%%%%%%%%%%%%%%%%%%%%%%%%%%%Figure Begin
\begin{figure}[htbp]
\centering
\includegraphics[width=\linewidth]{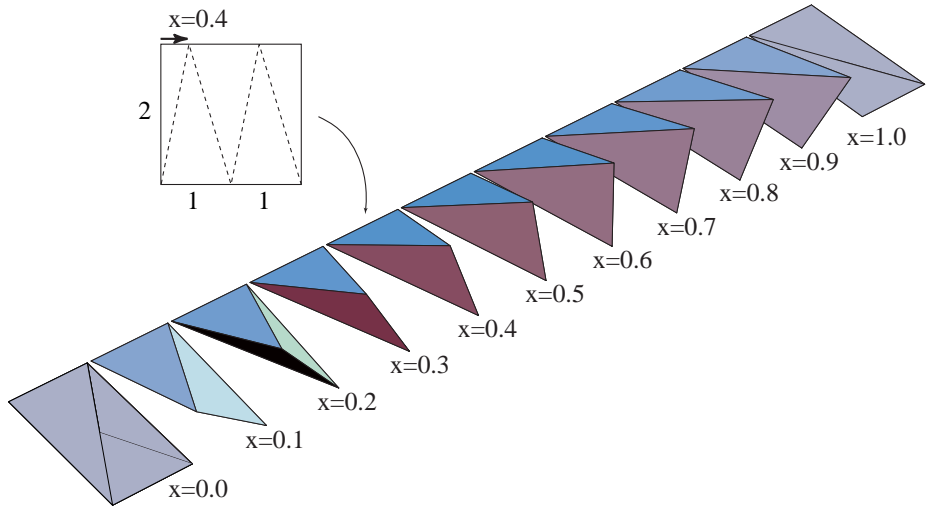}
\caption{Tetrahedra formed by folding a rectangle.}
\figlab{tetra}
\end{figure}
%%%%%%%%%%%%%%%%%%%%%%%%%%%%%%%%%Figure End

\subsection{Congruence: Unfolding}
\seclab{Congruence.Unfolding}

Although it is a long-standing open problem to determine
whether every polytope may be cut {\em along polytope edges\/}~\label{edge-unfolding} 
and unfolded to a polygon, 
without the ``along edges'' restriction it is easy
to see
that every polytope may be cut open to a continuum of 
noncongruent polygons.
To avoid trivial zigzaging of the cuts,
it makes sense to restrict the cuts to be 
{\em geodesics}, which unfold (or ``develop'') to straight lines,
or restrict even further to {\em shortest paths}, 
geodesics which are in addition
shortest paths between their endpoints.
Still this holds:

\begin{lemma}
Every polytope $Q$ may be cut via shortest paths
to unfold to a nondenumerably infinite number of noncongruent polygons.
\lemlab{star}
\end{lemma}
\full{
\begin{pf} {\em (Sketch)}
This may be accomplished via the star-unfolding~\cite{aaos-supa-97},\label{star-unfolding}
which cuts along the shortest paths from a source point $s$ to every
vertex of $Q$.
For any point $p$ in the interior of a 
``ridge-free region'' of $\bQ$,
every $s$ in
a neighborhood of $p$ yields a distinct star-unfolding.
\end{pf}
}%full

\section{Combinatorial Equivalence}
\seclab{Combinatorial.Equivalence}
Although a natural counterpart to our geometric enumerations would
count combinatorially distinct polygons and polytopes, the former class 
is uninteresting and the latter class seems difficult to capture.\footnote{
	Some results for convex unfoldings were obtained by 
	Shephard~\cite{s-cpcn-75}.
}
Instead we focus on the {\em process\/} of folding and unfolding, and
ask:
\begin{enumerate}
\squeezelist
\item How many combinatorially different foldings of a polygon lead to a polytope?
\item How many combinatorially different cuttings of a polytope lead to polygon
unfoldings?
\end{enumerate}
It requires some care to define an appropriate notion of
``combinatorially different'' for both questions.

\subsection{Combinatorics: Folding}
\seclab{Combinatorics.Folding}
We capture the combinatorics of a polygon folding via its ``gluing tree.''
Let a polygon $P$ have vertices $v_1,\ldots,v_n$,
labeled counterclockwise,
and edge $e_i$, $i=1,\ldots,n$ the open segment of $\bP$
after $v_i$.
The {\em combinatorial gluing tree\/} $T_G$
is a labeled tree representing the
identification of $\bP$ with itself.
%%%%%%%%%%%%%%%%%%%%%%%%%%%%%%%%%Figure Begin
\begin{figure}[htbp]
\centering
\includegraphics[width=0.9\linewidth]{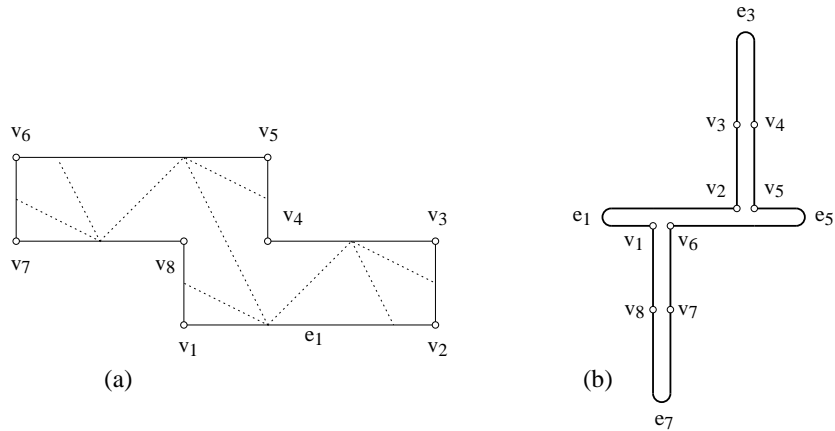}
\caption{(a) A polygon, with fold creases shown dotted;
(b) A gluing tree $T_G$ [folding away] corresponding to the crease pattern.}
\figlab{zed.tetra2}
\end{figure}
%%%%%%%%%%%%%%%%%%%%%%%%%%%%%%%%%Figure End
Any point of $\bP$ that is identified with more or less than
one other distinct point of $\bP$ becomes a node
of $T_G$, as well as any point to which a vertex is
glued.
(Note that this means there may be nodes of degree $2$.)
So every vertex of $P$ maps to a node of $T_G$;
each node is labeled with the set of all the
elements (vertices or edges) that are glued together there.
A point of $\bP$ in the interior of a polygon edge that glues only to itself, i.e., where a crease folds
the edge in two, we call a {\em fold point}.
\full{
Points $x$ and $y$ in Fig.~\figref{perim.halving} are fold points.
}%full
A fold-point correspond to a leaf of $T_G$,
and is labeled by the edge label only.
Every nonleaf node has at least one vertex label, and at most one edge label.
An example is shown in
Fig.~\figref{zed.tetra2}.
%\footnote{
%        Gluing trees may be drawn by folding up the polygon toward
%        the viewer, or folding the polygon
%        away (as in this figure).
%}
The polygon shown folds to a tetrahedron
by creasing as illustrated in (a).  All four tetrahedron
vertices are fold points.  The corresponding gluing tree
is shown in (b) of the figure.  The two interior nodes
of $T_G$ have labels $\{v_1,v_6,e_1\}$ and $\{v_2,v_5,e_5\}$.

We start with a characterization of the structure of gluing
trees, which will form the basis of our enumeration results.
\full{
Several combinatorial tree structures play a special
role, and to which we assign symbols:
\begin{enumerate}
\squeezelist
\item {\tt |}: a path.
\item {\tt Y}: a tree with a single degree-$3$ node.
\item {\tt I}: a tree with two degree-$3$ nodes connected by an edge
(e.g., Fig.~\figref{zed.tetra2}b).
\item {\tt +}: a tree with single degree-$4$ node.
\end{enumerate}

Next, define a {\em belt\/}
in a gluing tree to be a path between
two leaf fold points; for example,
between the $e_1$ and $e_3$ fold points in Fig.~\figref{zed.tetra2}b.
A belt is a {\em rolling belt\/}\label{rolling.belt}
if there is a nonzero-length
interval $I \subset e$ such that for every
$x \in I$, the belt folded at $x$ is an Aleksandrov
gluing.  
(A belt could instead have a finite number of distinct
gluings, perhaps just one.)

Our characterization shows that
gluing trees are fundamentally
discrete structures, with one or two rolling belts,
and two such belts only in very special circumstances:
\begin{theorem}
Gluing trees satisfy these properties:
\begin{enumerate}
\item At any gluing tree point of degree $d \neq 2$,
at most one point of $\bP$ in the interior of an edge may
be glued, i.e., at most one nonvertex may be glued there.
\item At most four leaves of the gluing tree can be fold points,
i.e., points in the interior of an edge of $\bP$.
The case of four fold-point leaves is only possible when
the tree has exactly four leaves, with the combinatorial
structure `{\tt +}' or `{\tt I}'.
\item A gluing tree can have at most two rolling belts.
\item A gluing tree with two rolling belts must
have the structure `{\tt I}', and result from folding
a polygon that can be viewed as a quadrilateral with two of
its opposite edges replaced by complimentary polygonal paths.
\end{enumerate}
\theolab{gluing.char}
\end{theorem}

Thus a generic gluing tree has one rolling  belt, with
trees hanging off it,
and one of those trees having a fold-point leaf,
as depicted in
Fig.~\figref{generic}.
}%full
\conf{
Due to lack of space in this abstract, we 
simply refer to Fig.~\figref{generic}.
}%conf
%%%%%%%%%%%%%%%%%%%%%%%%%%%%%%%%%Figure Begin
\begin{figure}[htbp]
\centering
\includegraphics[width=0.5\linewidth]{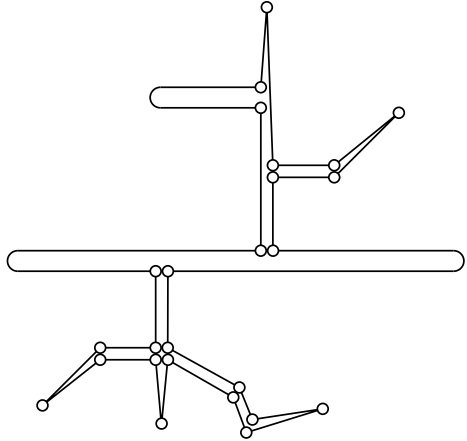}
\caption{A generic gluing tree: three fold-point leaves
(indicated by smooth arcs),
two forming a rolling belt.  Vertices indicated by open circles. }
\figlab{generic}
\end{figure}
%%%%%%%%%%%%%%%%%%%%%%%%%%%%%%%%%Figure End

We use this characterization to prove bounds on the number
of gluings.  First, a lower bound:
%%%%%%%%%%%%%%%%%%%%%%%%%%%%%%%%%Figure Begin
\begin{figure}[htbp]
\centering
\includegraphics[height=8.5cm]{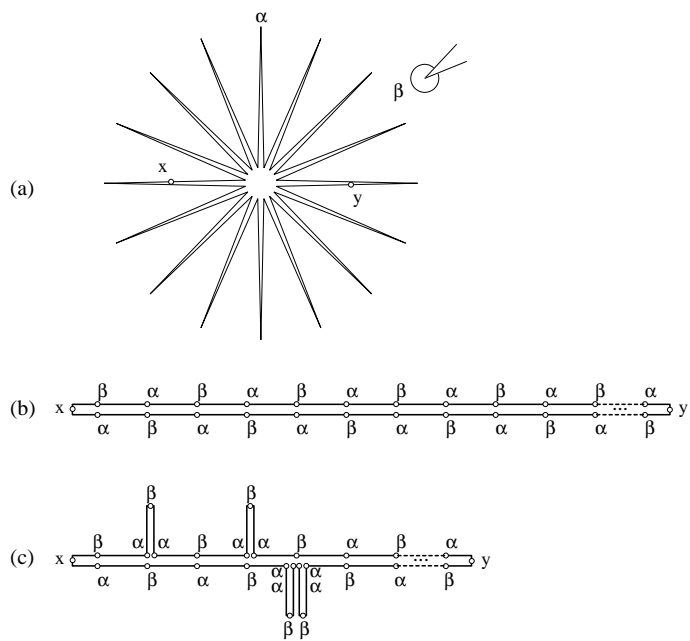}
\caption{(a) Star polygon $P$, $m=16$, $n'=32$, $n=34$.
(b) Base gluing tree.
(c) A gluing tree after several contractions.}
\figlab{star}
\end{figure}
%%%%%%%%%%%%%%%%%%%%%%%%%%%%%%%%%Figure End
\begin{theorem}
For any even $n$, there is a polygon $P$ of $n$ vertices
that has $2^{\Omega(n)}$ combinatorially distinct
Aleksandrov gluings.
\theolab{star}
\end{theorem}
\conf{
See Fig.~\figref{star}.
}%conf
\full{
\begin{pf} {\em (Sketch)}
The polygon $P$ is illustrated in Fig.~\figref{star}(a).
It is a centrally symmetric star, with $m$ vertices, $m$ even,
with a small convex angle $\a \approx 0$,
alternating with $m$ vertices
with large reflex angle $\b < 2\pi$.
All edges have the same (say, unit) length.
We call this an $m$-star.
We choose $\a$ small enough so that $m$ copies of $\a$ can
join with one of $\b$ and still be less than $2\pi$.
Now we add two vertices $x$ and $y$ at the midpoints of edges,
symmetrically placed so that $y$ is half the perimeter around
$\bP$ from $x$. 
Let $n=n'+2$ be the total number of vertices of $P$.

The ``base'' gluing tree is illustrated in
Fig.~\figref{star}(b).
$x$ and $y$ are fold vertices of the gluing.
Otherwise, each $\a$ is matched with a $\b$.
Because all edge lengths are the same, and because
$\a+\b < 2\pi$, %by Eq.~\eqref{a+b},
this path is an Aleksandrov gluing.
We label it $T_{00\cdots0,00\cdots0}$,
where $m/2$ zeros $00\cdots0$ represent the top chain, and another $m/2$
zeros represent the bottom chain.

The other gluing trees are obtained via ``contractions''
of the base tree.
A {\em contraction\/} makes any particular $\b$-vertex
not adjacent to $x$ or $y$
a leaf of the tree by gluing its two adjacent $\a$-vertices
together.  Label a $\b$-vertex $0$ or $1$ depending
on whether it is uncontracted or contracted
respectively.  Then a series of contractions can be identified
with a binary string.
For example, Fig.~\figref{star}(c) displays the
tree $T_{010100\cdots,00110\cdots0}$.
%Note that $k$ adjacent contractions result in $2k$
%$\a$-vertices glued together.

We can bound the number of Aleksandrov gluings resulting from
these contractions by 
$\Omega(2^{m/2-1}) = %\Omega(2^{(n-6)/4}) = 
2^{\Omega(n)}$.
\end{pf}
}%full

%%%%%%%%%%%%%%%%%%%%%%%%%%%%%%%%%Figure Begin
%\begin{wrapfigure}{l}[2mm]{30mm}
%\centering
%\includegraphics[height=0.2\textheight]{Latin.eps}
%\caption{Latin cross.}
%\figlab{Latin}
%\end{wrapfigure}
%%%%%%%%%%%%%%%%%%%%%%%%%%%%%%%%%Figure End
It may not be surprising that some polygons have many foldings,
but it is perhaps less intuitive that even simple polygonal
shapes have many gluings.
For example, our enumeration program finds that an equilateral
triangle has 19 gluings and a square 43 gluings.
Of the latter, 10 foldings are distinct when symmetries are
removed: several flat shapes, four tetrahedra, a hexahedron, and a
continuum of octahedra.
%Even more interesting,
Hirata~\cite{h-pc-00} has shown that the Latin square,
%(Fig.~\figref{Latin}),
whose study we initiated
in~\cite{lo-wcpfp-96}, has 85 distinct gluings.
These lead to 21 distinct shapes,
%In Erik's list:
%1,2,3,5,11,12,18,19,20,24,30,31,32,34,43,52,53,61,64,65,67
including
several
flat quadrangles,
tetrahedra,
hexahedra (including a cube),
octahedra,
and a pentahedron.\footnote{
        \url{http://daisy.uwaterloo.ca/~eddemain/aleksandrov/cross/} .
}
%~\cite{ddlo-85flc-00}.
\full{
Fig.~\figref{65} shows crease patterns for two gluings.
%%%%%%%%%%%%%%%%%%%%%%%%%%%%%%%%%Figure Begin
\begin{figure}[htbp]
\begin{minipage}[b]{0.5\linewidth}
\centering
\includegraphics[width=0.5\linewidth]{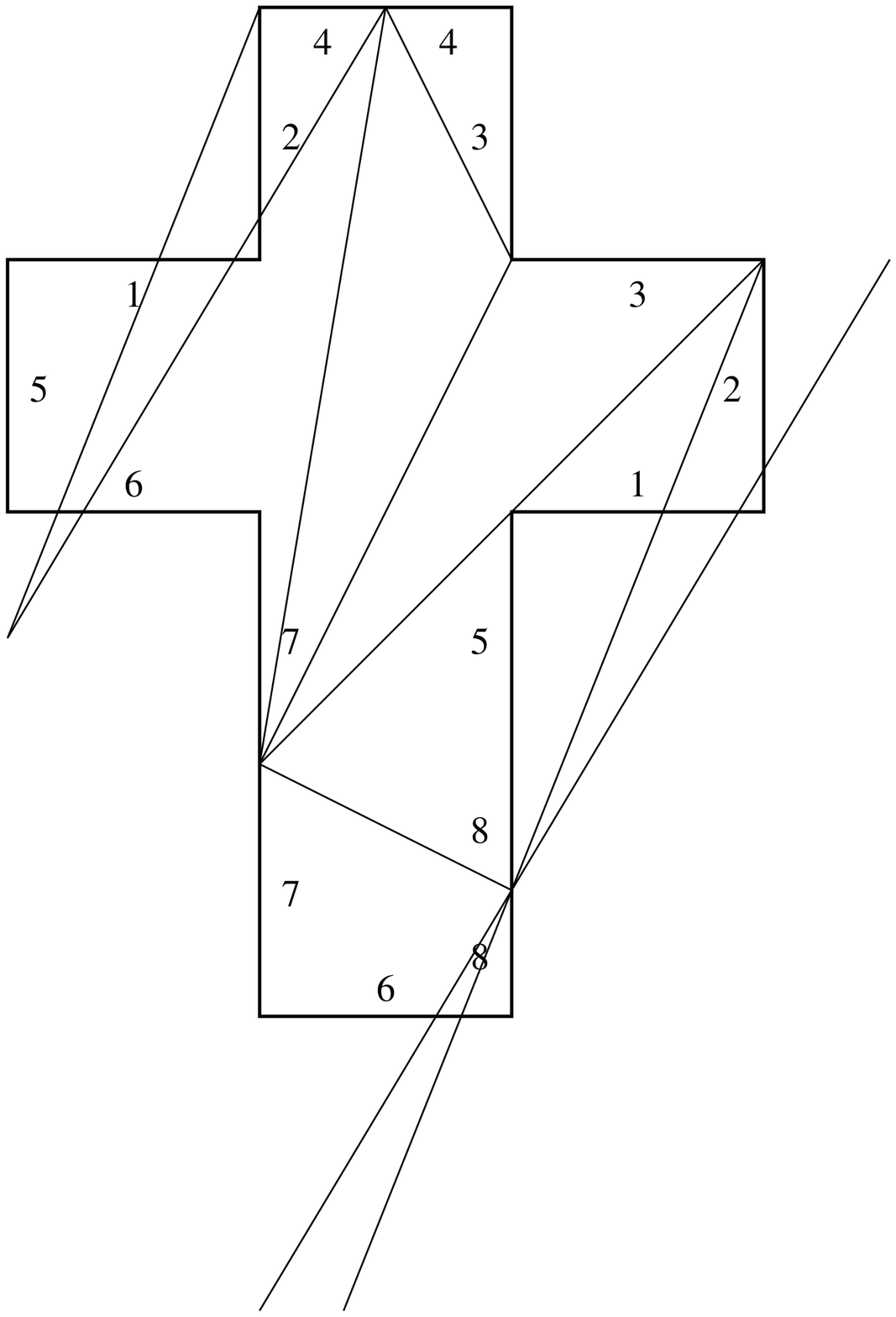}
\end{minipage}%
\begin{minipage}[b]{0.5\linewidth}
\centering
\includegraphics[width=4cm]{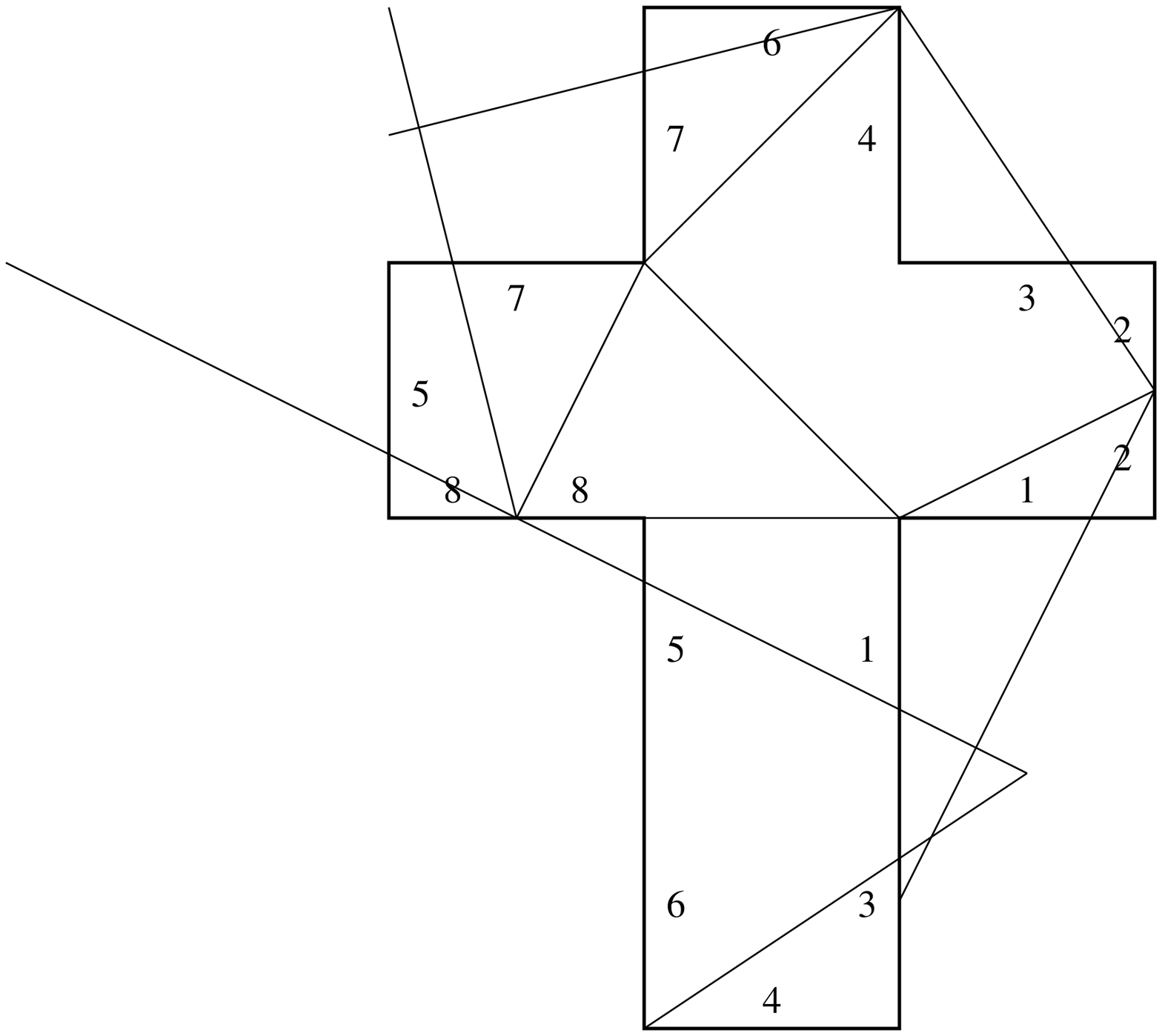}
\end{minipage}%
\caption{Construction lines for creases to fold 
a Latin cross 
to a hexahedron (left) and
a pryramid with a quadrangular base (right).}
\figlab{65}
\end{figure}
%%%%%%%%%%%%%%%%%%%%%%%%%%%%%%%%%Figure End
}%full

We may obtain an upper bound in terms of the number
of leaves $\l$ of the gluing tree:
\begin{theorem}
The number of gluing trees with $\l$ leaves for a polygon $P$
with $n$ vertices is $O(n^{2\l-2})$.
\theolab{gluing.upper}
\end{theorem}

\noindent
This bound is useful when the number of leaves is bounded,
which is the case, for example, with convex polygons.
\full{
The characterization of Theorem~\theoref{gluing.char} can be
tightened in this case:
}%full
\conf{
The characterization indicated in Fig.~\figref{generic} can be
tightened in this case:
}%conf

\begin{lemma}
For convex polygons, the gluing tree
$T_G$ has one of
these combinatorial
structures: 
when $n \neq 4$, either `{\tt |}' or `{\tt Y}';
when $n = 4$, in addition `{\tt I}' and `{\tt +}' are possible.%
\footnote{
	This lemma is largely due to Shephard~\cite{s-cpcn-75}.
}
\lemlab{char.convex}
\end{lemma}
This leads to a tighter bound for convex polygons:
\begin{theorem}
A convex polygon $P$ of $n$ vertices folds to at most
$O(n^3)$ different gluing trees.
Some convex polygons have $\Omega(n^2)$ gluings.
\theolab{count.gluing.trees}
\end{theorem}
We leave open the task of closing the gap between quadratic and cubic.

\subsection{Combinatorics: Unfolding}
\seclab{Combinatorics.Unfolding}
Finding the ``right'' way to count unfoldings is more delicate.
We start by defining cut trees, which then form the basis
of our enumerations.
It will be useful to distinguish between
a {\em geometric\/} tree $\T$ composed of a union of line
segments, and the more familiar {\em combinatorial\/} tree $T$
of nodes and arcs.
A {\em geometric cut tree\/} $\T_C$ 
for a polytope $Q$ is a tree drawn on $\bQ$,
with each arc a polygonal path,
which leads to a polygon unfolding when the surface is cut along $\T$,
i.e., flattening $Q \setminus \T$ to a plane.

\begin{lemma}
If a polygon $P$ folds to a polytope $Q$, $\bP$ maps
to a tree $\T_C \subset \bQ$, the geometric cut tree, with
the following properties:
\begin{enumerate}
\squeezelist
\item $\T_C$ is a tree.
\item $\T_C$ spans the vertices of $Q$.
\item Every leaf of $\T_C$ is at a vertex of $Q$.
\item
A point of $\T_C$ of
degree $d$ (i.e., one with $d$ incident segments)
corresponds to exactly $d$ points of $\bP$.
Thus a leaf corresponds to a unique point
of $\bP$.
\item Each arc of $T_C$ is a polygonal path on $Q$.
\end{enumerate}
\lemlab{cut.tree}
\end{lemma}

\full{
\noindent
There are several options in defining the combinatorial tree $T_C$
for a geometric $\T_C$:
\begin{enumerate}
\squeezelist
\item Make every segment of $\T_C$ an arc of $T_C$.
Although this is natural, 
allowing an arbitrarily
complicated polygonal path between any two polytope vertices
leads to an infinite number
of different cut trees for any polytope.
%, leading to different combinatorial trees.
\item Make every point where a path of $\T_C$ crosses an edge
of the polytope a node of $T_C$.  This again leads to trivially
infinite numbers of cut trees when a path of $\T_C$ zigzags
back and forth over an edge of $Q$.
\item Exclude this possibility by forcing the paths between
polytope vertices to be
geodesics, and again make polytope edge crossings nodes of $T_C$.
This excludes many interesting cut trees 
%---all those where a polygon vertex is glued to a point with angle sum $2\pi$---
and destroys symmetry between $T_C$ and $T_G$.
\item Make every maximal path of $\T_C$ consisting only of
degree-$2$ points a single arc of $T_C$.  
A consequence is that
polytope vertices in the interior
of such a path disappear from $T_C$.
\end{enumerate}

Threading between these possibilities, we define
the {\em combinatorial cut tree\/} $T_C$ corresponding
to a geometric cut tree $\T_C$ as
the labeled graph with a node (not necessarily labeled)
for each point of $\T_C$ with degree not equal to $2$,
and a labeled node for each point of $\T_C$ that corresponds
to a vertex of $Q$ (labeled by the vertex label);
arcs are determined by the polygonal paths of $\T_C$
connecting these nodes.
An example is shown in
Fig.~\figref{cut.tree}.  Note that not every node
of the tree is labeled, but
every polytope vertex label is used at some node.
All degree-$2$ nodes are labeled.
%%%%%%%%%%%%%%%%%%%%%%%%%%%%%%%%%Figure Begin
\begin{figure}[htbp]
\centering
\includegraphics[width=0.6\linewidth]{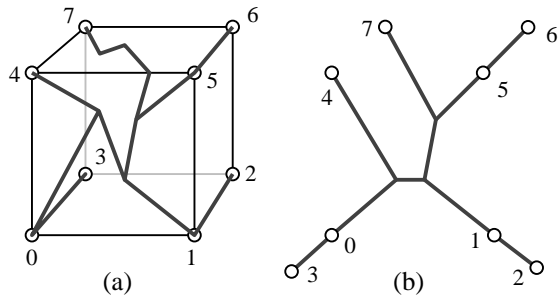}
\caption{(a) Geometric cut tree $\T_C$ on the surface of a cube;
(b) The corresponding combinatorial cut tree $T_C$.}
\figlab{cut.tree}
\end{figure}
%%%%%%%%%%%%%%%%%%%%%%%%%%%%%%%%%Figure End

This definition has the consequence that, if all
degree-$2$ nodes are removed by contraction,
$T_C$ is isomorphic to the corresponding gluing tree $T_G$.
Although the definition avoids some of the listed pitfalls,
it does have the
undesirable consequence of counting different geodesics
on $\bQ$ between two polytope vertices as the same arc of $T_C$
even if one spirals around the polytope twice and the other once
(or not at all).

Before turning to enumeration bounds, we make this straightforward observation:
Every polytope admits at least the $n$ cut trees
provided by the 
star-unfolding~\cite{ao-nsu-92}, %(p.\pageref{star-unfolding}), 
one with each
vertex as source.
So in particular, every polytope has at least one unfolding to a simple polygon,
in contrast to the corresponding
open question for edge-unfoldings (p.\pageref{edge-unfolding}).
}%full

Our main result here is that some polytopes have an exponential number
of unfoldings:
\begin{theorem}
There is a polytope $Q$ of $n$ vertices
that may be cut open with exponentially many {\rm ($2^{\Omega(n)}$)}
combinatorially distinct
cut trees,
which unfold to exponentially many geometrically distinct simple polygons.
\theolab{volcano}
\end{theorem}
\conf{
Two different cuttings of $Q$ are illustrated in
Figs.~\figref{volcano.1} and~\figref{volcano.2}.
}%conf
\full{
\begin{pf} {\em (Sketch)}
$Q$ is a truncated cone:
%Fig.~\figref{volcano.0}:
%%%%%%%%%%%%%%%%%%%%%%%%%%%%%%%%%%Figure Begin
%\begin{figure}[htbp]
%\centering
%\includegraphics[width=0.5\linewidth]{volcano.0.eps}
%\caption{
%Polytope $Q$.
%}
%\figlab{volcano.0}
%\end{figure}
%%%%%%%%%%%%%%%%%%%%%%%%%%%%%%%%%%Figure End
the hull of
two regular $n$-gons of different radii, lying in parallel planes
and similarly oriented.
Two different cuttings are illustrated in 
Figs.~\figref{volcano.1} and~\figref{volcano.2}.
%We call this the {\em volcano\/} example.
}%full
%%%%%%%%%%%%%%%%%%%%%%%%%%%%%%%%%Figure Begin
\begin{figure}[htbp]
\centering
\includegraphics[width=0.9\linewidth]{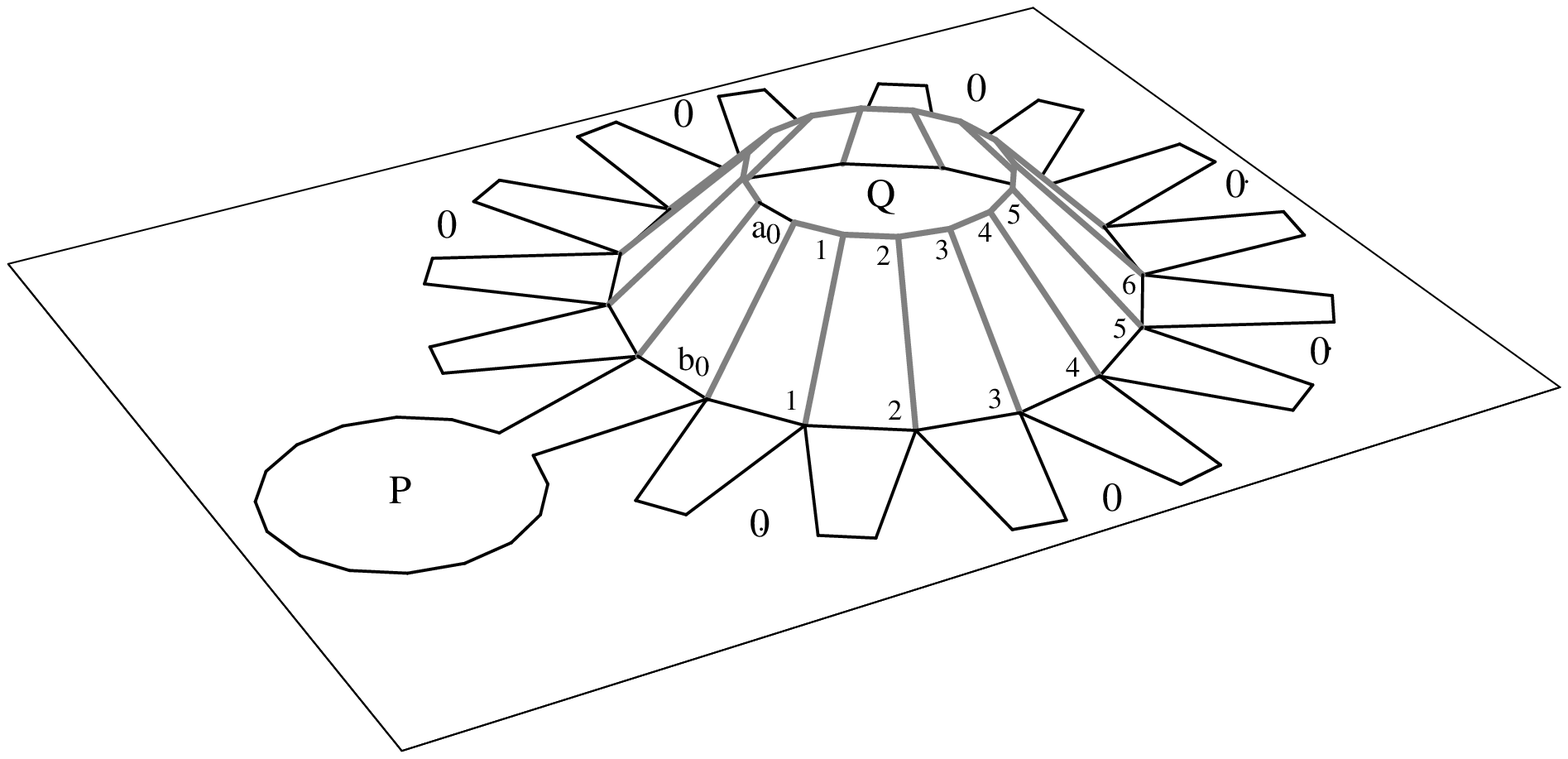}
\caption{
Unfolding via shaded cut tree $T_{0000000}$.
}
\figlab{volcano.1}
%\end{figure}
\vspace{5mm}
%\begin{figure}[htbp]
\centering
\includegraphics[width=0.9\linewidth]{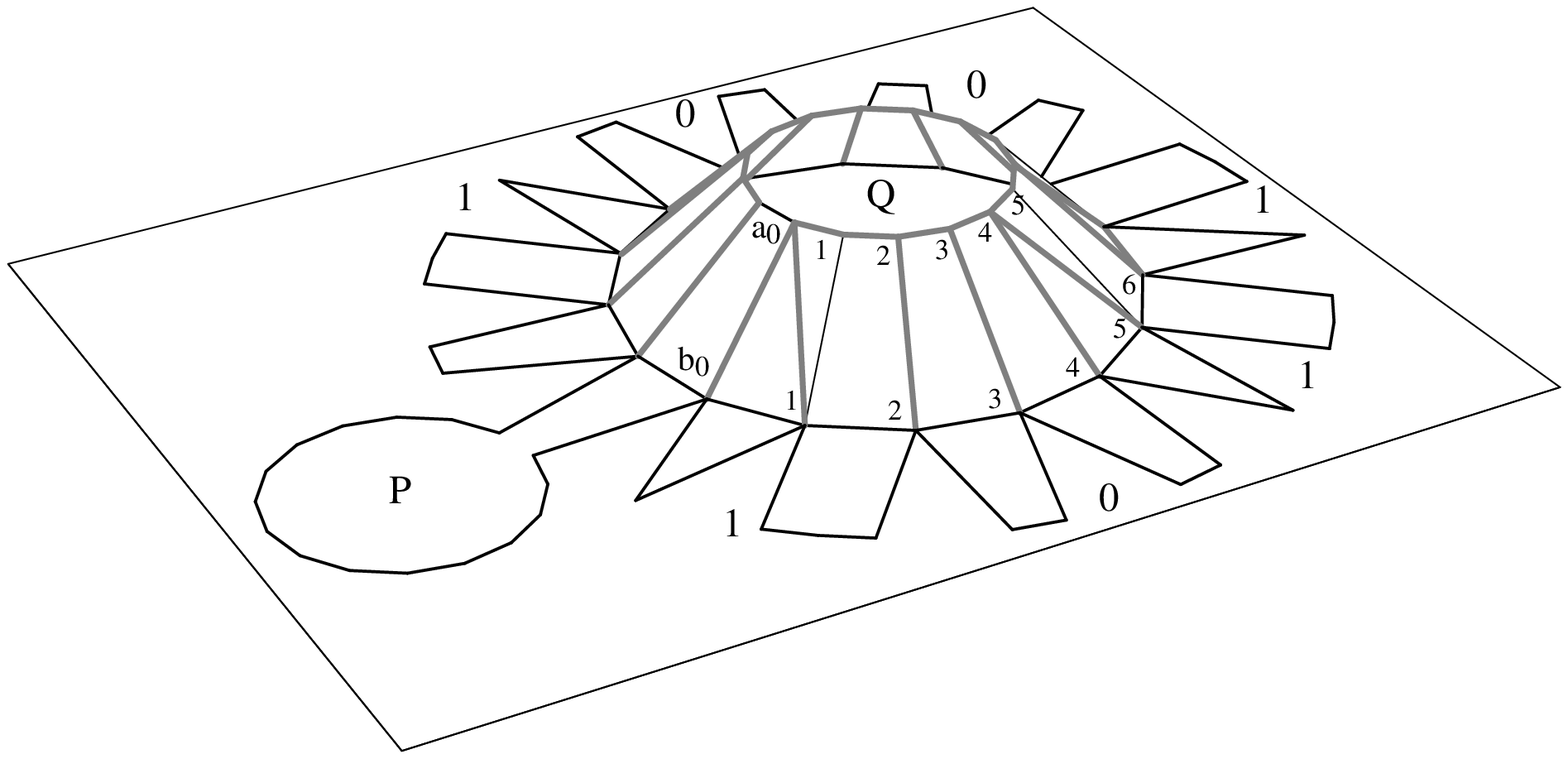}
\caption{
Unfolding via shaded cut tree $T_{1001101}$.
}
\figlab{volcano.2}
\end{figure}
%%%%%%%%%%%%%%%%%%%%%%%%%%%%%%%%%Figure End
\full{
The ``base'' cut tree, which we notate as $T_{0000000}$,
unfolds $Q$ as shown in Fig.~\figref{volcano.1}.
Define a cut tree $T_{m_{(n-1)/2} \cdots m_2 m_1 m_0}$,
where $m_i$ are the digits of a binary number
of $n/2 - 1$ bits,
as an alteration of the base tree $T_{0\cdots0}$
illustrated by $T_{1001101}$ shown in Fig.~\figref{volcano.2}.
There are $2^{n/2 - 1} = 2^{\Omega(n)}$ cut trees,
and it is not difficult to show that each leads to a distinct
simple polygon unfolding.
\end{pf}
}%full

\noindent
We conjecture that 
there is a polytope with an exponential number of
convex unfoldings, i.e., those that result in convex polygons.

Our upper bound relies on bounds on the number of
spanning trees of triangulated planar graphs:
\begin{theorem}
The maximum number of combinatorially distinct edge-unfolding cut trees of a polytope of $n$
vertices is $2^{O(n)}$,
and the maximum number of (arbitrary) combinatorially distinct cut trees
is $2^{O(n^2)}$.
\end{theorem}

\section{Open Problems}
Some of the most interesting open questions in this area are algorithmic:
\begin{enumerate}
\squeezelist
\item Given an Aleksandrov gluing, compute
the 3D structure of the polytope.  
This is an algorithmic version of Aleksandrov's theorem,
for which only a ``finite'' algorithm is known.\footnote{
	I.~K.~Sabitov, Oberwolfach presentation, May 2000.
	See also \url{http://www.cms.math.ca/CMS/Events/winter98/w98-abs/node46.html}.
}
This problem is closely related to the following problem,
which may be easier because of the additional information:
\item Given an Aleksandrov gluing and the unique crease pattern
for the folding
\full{
(Cf.~Fig.~\figref{65}),
}%full
compute
the 3D structure of the polytope.
This can be viewed as an algorithmic version of Cauchy's rigidity theorem.
\item How difficult is it to determine whether a given polytope
has a convex unfolding? (Cf.~Lemma~\lemref{char.convex}.)
\item How difficult is it to determine whether a given polygon
may be folded to a polytope?
(Cf.~Fig.~\figref{unf}.)
We have an exponential-time algorithm, but a polynomial time algorithm
is known only for edge-to-edge gluings~\cite{lo-wcpfp-96}.
\end{enumerate}

%%%%%%%%%%%%%%%%%%%BIB
% Make the bibliography
\small
\bibliographystyle{alpha}
\bibliography{enum}
\end{document}